\documentclass[pre,aps,twocolumn]{revtex4}

\usepackage{graphicx}
\usepackage{amsmath,empheq}
\usepackage{amssymb}
\usepackage{ifthen}
\usepackage{booktabs}
\usepackage{graphicx}
\usepackage{dcolumn}
\usepackage{bm}
\usepackage{longtable}
\usepackage{gensymb}

\newcommand{\bb}{\begin{equation}}
\newcommand{\ee}{\end{equation}}
\newcommand{\ba}{\begin{eqnarray*}}
\newcommand{\ea}{\end{eqnarray*}}

\begin{document}

\title{Edge Contact Angle, Capillary Condensation, and Meniscus Depinning}

\author{Alexandr \surname{Malijevsk\'y}}
\affiliation{ {Department of Physical Chemistry, University of Chemical Technology Prague, Praha 6, 166 28, Czech Republic;}
 {The Czech Academy of Sciences, Institute of Chemical Process Fundamentals,  Department of Molecular Modelling, 165 02 Prague, Czech Republic}}
 \author{Andrew O. \surname{Parry}}
\affiliation{Department of Mathematics, Imperial College London, London SW7 2BZ, UK}

\begin{abstract}
We study the phase equilibria of a fluid confined in an open capillary slit formed when a wall of finite length $H$ is brought a distance $L$ away
from a second macroscopic surface. This system shows rich phase equilibria arising from the competition between two different types of capillary
condensation, corner filling and meniscus depinning transitions depending on the value of the aspect ratio $a=L/H$. For long capillaries, with
$a<2/\pi$, the condensation is of type I involving menisci which are pinned at the top edges at the ends of the capillary characterized by an edge
contact angle. For intermediate capillaries, with $2/\pi<a<1$, depending on the value of the contact angle the condensation may be of type I or of
type II, in which the menisci overspill into the reservoir and there is no pinning. For short capillaries, with $a>1$, condensation is always of type
II. In all regimes, capillary condensation is completely suppressed for sufficiently large contact angles. We show that there is an additional
continuous phase transition in the condensed liquid-like phase, associated with the depinning of each meniscus as they round the upper open edges of
the slit. Finite-size scaling predictions are developed for these transitions and phase boundaries which connect with the fluctuation theories of
wetting and filling transitions. We test several of our predictions using a fully microscopic Density Functional Theory which allows us to study the
two types of capillary condensation and its suppression at the molecular level.
\end{abstract}

\maketitle

The contact angle $\theta$ is central to the study of fluid adsorption and plays a crucial role in a number of surface phase transitions
 where it specifies the phase boundary \cite{rowlin, dietrich, bonn}. For example, it vanishes at a wetting transition \cite{cahn, ebner} and also
determines that a right-angle corner is filled by liquid when $\theta<\pi/4$ \cite{hauge, rejmer}. It also appears in the macroscopic Kelvin equation
for the pressure shift from saturation, $\delta p_{\rm cc}=2\gamma \cos \theta/L$,  where $\gamma$ is the interfacial tension, at which a vapour
confined between two identical plates separated by a distance $L$, condenses to liquid \cite{gregg, evans85}. If the walls are materially different,
this generalizes immediately to $\delta p_{\rm cc}=\gamma (\cos\theta_1+\cos\theta_2)/L$, with $\theta_1$ and $\theta_2$ the corresponding contact
angles. There are well known cases where the equilibrium value of $\theta$ is modified by interfacial pinning, e.g., on rough surfaces where the
modification is described by Wenzel's law \cite{degennes} and is believed to underline the phenomena of contact angle hysteresis \cite{quere}.
Interfacial pinning is also important for condensation in open capillaries. In particular, recent studies of fluid equilibria in a slit of finite
length have highlighted the role played by an {\it{edge contact angle}} which characterizes the menisci pinned at the ends, and which replaces
$\theta$ in the Kelvin equation \cite{fin_slit, laska20, geim20}.

\begin{figure}
\includegraphics[width=4.cm]{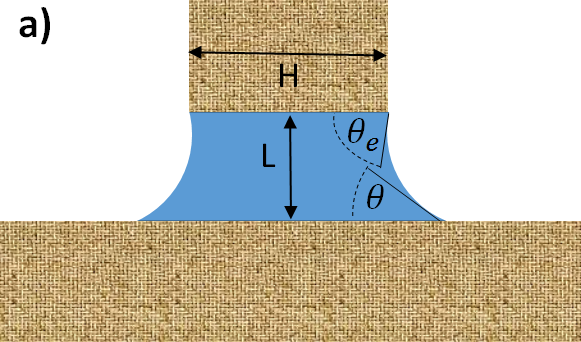} \hspace*{0.1cm} \includegraphics[width=4.2cm]{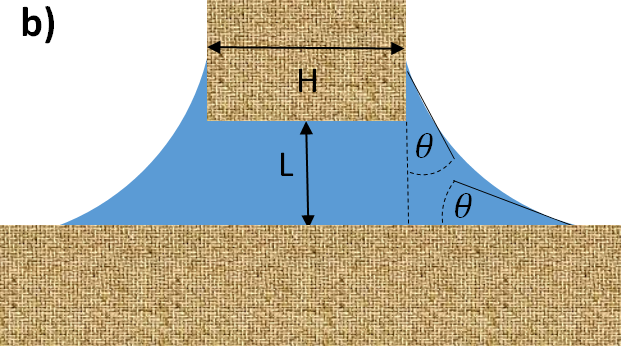}

\caption{Schematic illustration of two possible capillary liquid (CL) configurations in which the menisci a) are pinned at the top corners with edge
contact angle $\theta_e$, b) spill out of the slit meeting the walls with Young's contact angle $\theta$.} \label{sketch_hinf}
\end{figure}

In this paper, we study a fluid confined in a slit formed when a wall of length $H$ is brought near a second, infinite surface. We show that due to a
combination of interfacial phenomena, the phase behaviour is extremely rich. In particular, there are two types of capillary condensation as well as
a continuous interfacial transition which has not been identified previously. At all these phase boundaries, an equilibrium edge contact angle plays
a crucial role. The possible phase diagrams fall into three universal classes depending on the aspect ratio $a=L/H$. The rounding of these
transitions, occurring at the mesoscopic scale, is linked to the fluctuation theories of wetting and filling transitions although several aspects are
observable at the truly microscopic level.

\begin{figure*}
\includegraphics[width=4.8cm]{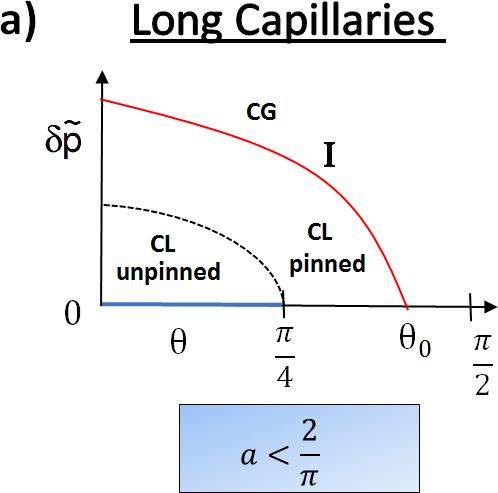} \hspace*{0.3cm} \includegraphics[width=5.5cm]{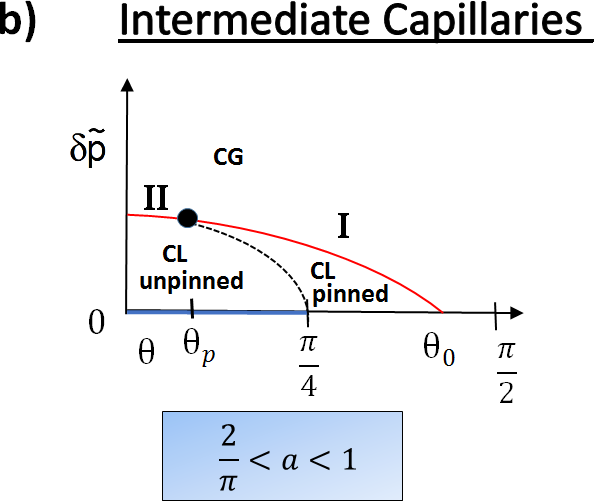} \hspace*{0.3cm}  \includegraphics[width=4.8cm]{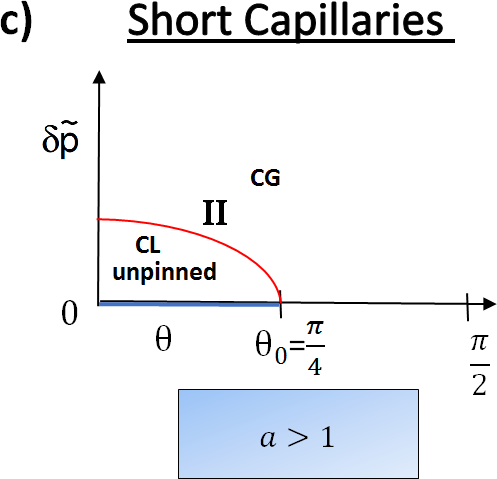}
\caption{Macroscopic phase diagrams for long, intermediate and short capillaries showing the location of type I and type II condensation (red line),
the meniscus depinning transition (dashed line) and the suppression of capillary condensation for different values of the aspect ratio. Here $\tilde
\delta p= L \delta p/(2\gamma)$ is a dimensionless measure of the deviation from bulk saturation.} \label{pd_theta-dp}
\end{figure*}

We begin with macroscopics and suppose that our system is in contact with a reservoir of gas at pressure $p$ at a temperature $T$  below that of the
bulk critical point $T_c$. Translational invariance is assumed along the slit and gravity is neglected which is valid, for molecular fluids, provided
$L$ is sub $mm$ \cite{degennes}. We anticipate that as $p$ is increased, the fluid inside the capillary condenses to liquid before bulk saturation,
$p_{\rm sat}$, is reached. The liquid-like phase is characterized by two circular menisci, which must be located near the ends. Since the bottom wall
is infinite, the menisci meet it at Young's contact angle $\theta$ but there are two possibilities for the upper part of each menisci. For example,
they may be pinned at the edges, making an angle $\theta_e$ w.r.t. the horizontal (see Fig.~\ref{sketch_hinf}a). This edge contact angle is pressure
dependent for any capillary-liquid phase (CL) but takes a specific value $\theta_e^{\rm cc}$ at the first-order transition where it coexists with the
capillary-gas phase (CG); we refer to this as type I capillary condensation. Alternatively, the menisci may be unpinned, sitting outside the open
ends touching the walls with the contact angle $\theta$ (see Fig.~\ref{sketch_hinf}b); we refer to this as type II capillary condensation.

The edge contact angle of any CL is determined geometrically by $L/R=\cos\theta+\cos\theta_e$ where $R=\gamma/\delta p$ is the Laplace radius and
$\delta p$ is the pressure difference across the meniscus which is approximately the deviation from $p_{\rm sat}$. The maximum value of the edge
contact angle is $\theta_e^{\rm max}=\theta+\pi/2$,  which is when the upper part of the menisci meet the vertical walls at Young's contact angle and
are therefore unpinned. Balancing the grand potentials of the CG and CL phases taking into account only the volume and surface contributions shows
that type I capillary condensation occurs when
 \begin{equation}
\delta {p}_{\rm cc}^I=\frac{\gamma}{L}(\cos\theta+\cos\theta_e^{\rm cc})\,, \label{pcc_long}
 \end{equation}
 where $\theta_e^{\rm cc}$ is determined implicitly from
 \begin{equation}
\cos^2\theta=\cos^2\theta_e^{\rm cc} +a\,\frac{\pi-\theta-\theta_e^{\rm cc}+\sin(\theta+\theta_e^{\rm cc})}{1+a\tan\left(\frac{\theta_e^{\rm
cc}-\theta}{2}\right)}\,. \label{cosplus}
\end{equation}
The modified Kelvin equation (\ref{pcc_long}) is therefore of the same form as that for an infinite slit with two materially distinct walls. When the
slit is infinitely long we recover the standard Kelvin equation, since $\theta_e^{\rm cc}=\theta$. As we shorten the capillary, the value of
$\theta_e^{\rm cc}$ increases and the condensation occurs closer to $p_{\rm sat}$. The loci of type I condensation terminates in one of two ways. It
ends if $\theta_e^{\rm cc}=\theta_e^{\rm max}$ when it becomes of type II and no longer involves pinned menisci. Again, balancing the macroscopic
contributions to the grand potentials of the CG and CL phases determines that this type of condensation occurs at the pressure shift
 \begin{equation}
\delta {p}_{\rm cc}^{II}=\frac{2\gamma}{L}\frac{a\left[ \cos\theta-\sin\theta+\left(\theta-\frac{\pi}{4}\right)\sec\theta\right]}{a-1+\sqrt{1+a^2-2a
\left(\frac{\pi}{4}\sec^2\theta-\tan\theta\right)}}\,. \label{pshort}
 \end{equation}
The numerator is positive only for $\theta<\pi/4$ implying that type II condensation can only occur in the complete corner filling regime.
Alternatively, type I condensation ends when $\theta_e^{\rm cc}=\pi-\theta$, for which $\delta p_{\rm cc}^I=0$. From Eq.~(\ref{cosplus}) it follows
that this occurs when the aspect ratio is $a_0=\cot\theta$. We find it remarkable that the capillary condensation at this terminus of type I
condensation mimics the phase separation in an infinite slit where the walls are materially different with opposing wetting properties, i.e.,
$\theta_2=\pi-\theta_1$ \cite{parry90}. Capillary condensation is suppressed for shorter capillaries.

We can summarize these results using simple phase diagrams. We begin by classifying the capillaries according to their aspect ratio:

\begin{figure*}
\includegraphics[width=6cm]{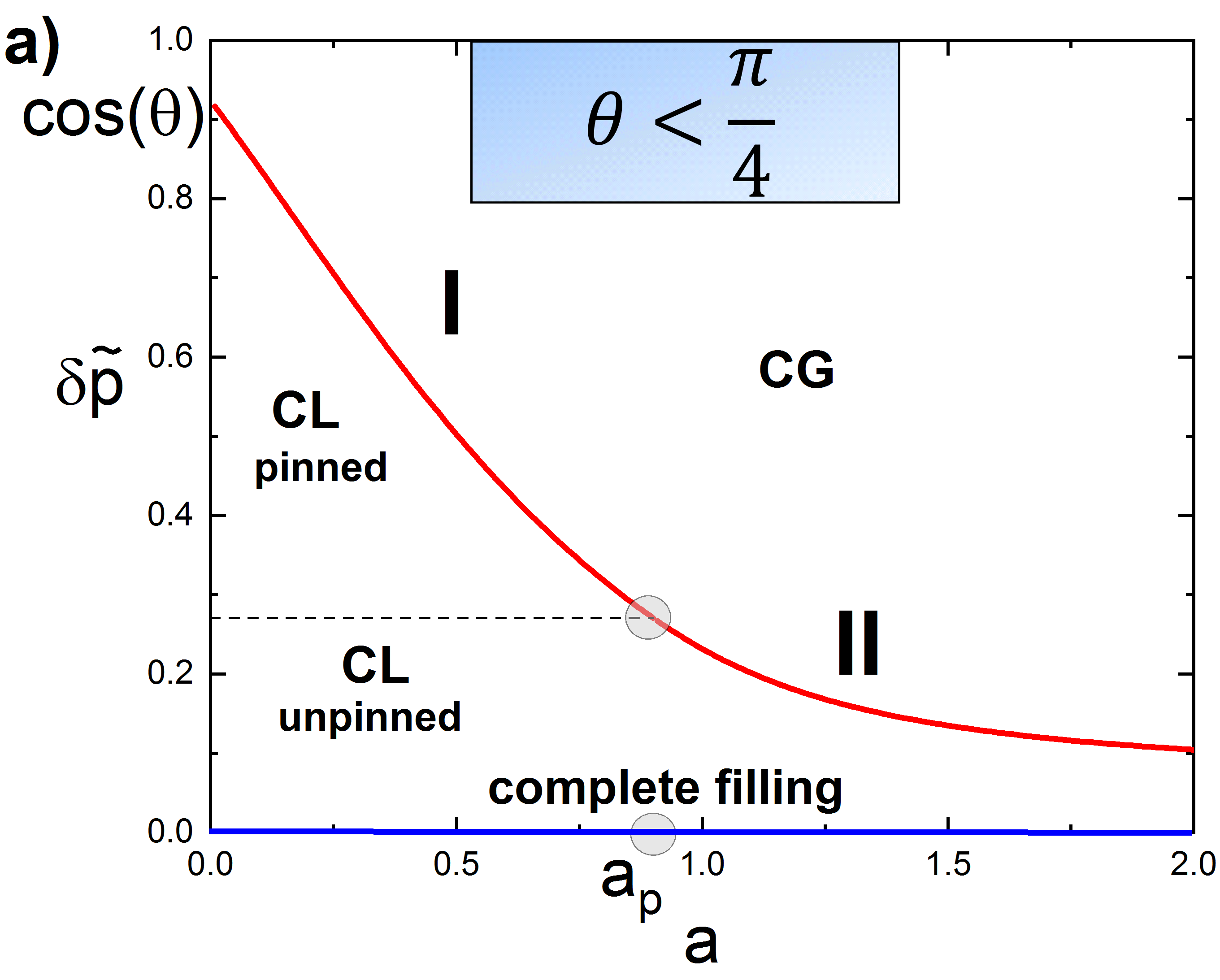} \hspace*{1cm} \includegraphics[width=6cm]{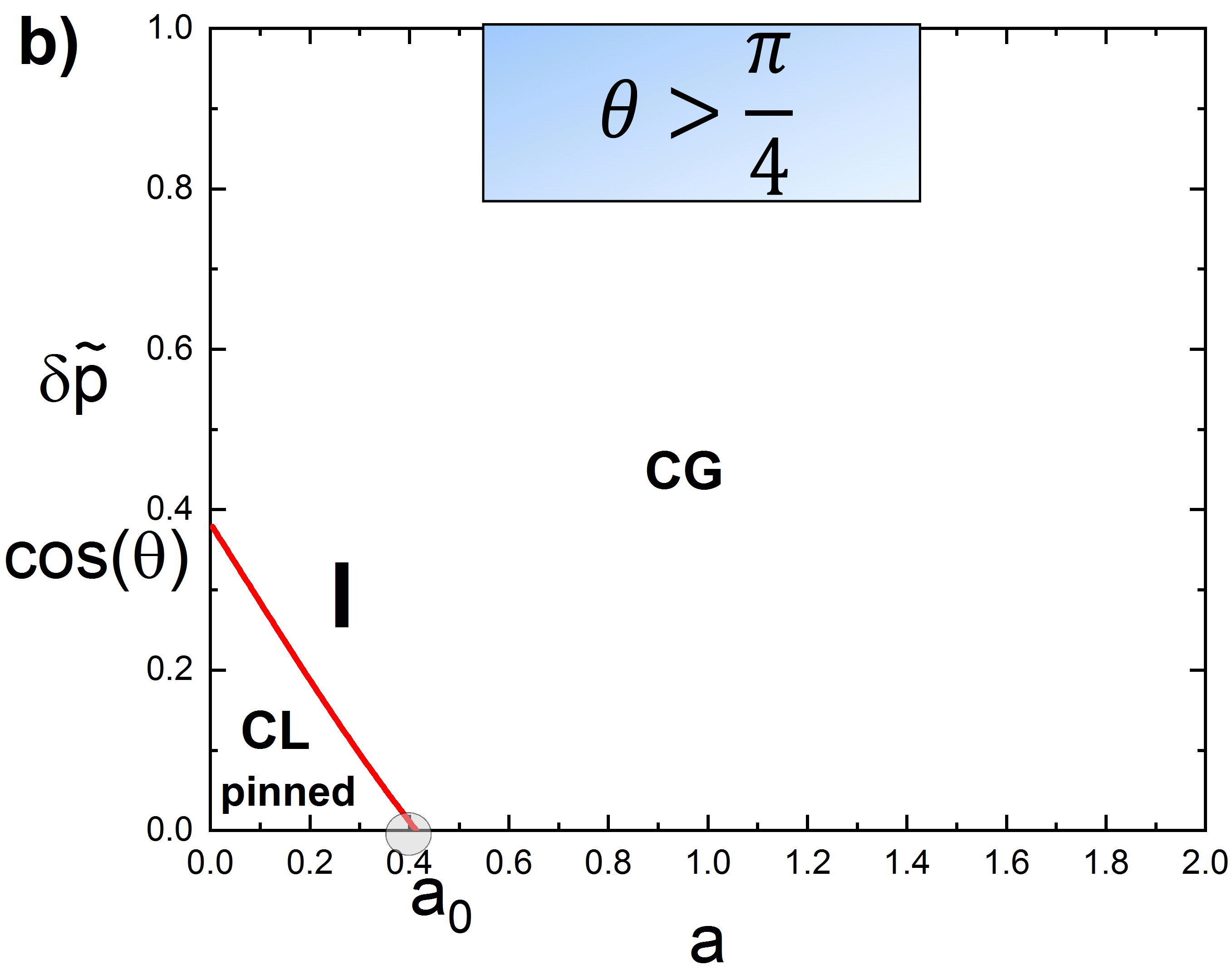}
\caption{Macroscopic phase diagrams for complete and partial corner filling. For $\theta<\pi/4$ (a) both type I and type II condensation (red line)
occur in addition to meniscus depinning (dashed line) and complete corner filling (blue line). The type I/II crossover occurs at $a_p=\cos
(2\theta)/(\pi/2-2 \theta)$. As $\theta$ increases to $\pi/4$ the lines of meniscus depinning and type II condensation merge into the saturation
curve and disappear. For $\theta>\pi/4$ (b) only type I condensation exists and is suppressed when the aspect ratio is larger than $a_0=\cot
\theta$.} \label{dp-a}
\end{figure*}

{ \bf Long capillaries} (Fig.~2a). For $a<2/\pi$, only type I condensation occurs up to a maximum value of the contact angle $\theta_0=\cot^{-1}a$.
For $\theta\ge \theta_0$ the gas inside the slit and the surrounding reservoir both condense to liquid at $p_{\rm sat}$. In addition, there is a line
of meniscus depinning transitions when
 \bb
 \delta p_{\rm md}=\frac{\gamma(\cos\theta-\sin\theta)}{L}\,, \label{cf_pmd}
 \ee
extending to the corner filling phase boundary $\theta=\pi/4$. On increasing the pressure from $\delta p_{\rm cc}^I$ the edge contact angle increases
until it reaches $\theta_e=\theta_e^{\rm max}$, at which point the meniscus depins. This is shown as the dashed line in Fig.~\ref{pd_theta-dp}a and
separates the regimes where the upper parts of the menisci are pinned or unpinned. Meniscus depinning is a continuous phase transition which is
third-order for complete wetting and second-order for partial wetting.  For example, for complete wetting,  the third derivative of the grand
potential, $\partial^3\Omega/\partial R^3$, has a discontinuity $\gamma/L^2$ associated with a singularity in the adsorption $\Gamma_{\rm
sing}\propto(\delta p_{\rm md}-\delta p)^2$, which arises from the different qualitative structure of the meniscus in the pinned and the unpinned
regimes. Finally the (blue) line, $0<\theta<\pi/4$, at bulk saturation represents the line of complete corner filling. On approaching this line the
adsorption diverges as $\Gamma\propto R^2$ due to the continuous growth of two menisci which have spilled out into the right-angle corners at each
end of the slit.

{\bf Intermediate capillaries} (Fig.~2b). If $2/\pi<a<1$, then condensation is of type II for $0<\theta<\theta_p$ and type I for
$\theta_p<\theta<\theta_0$. The crossover occurs when $a=\cos (2\theta_p)/(\pi/2-2\theta_p)$, at which $\theta_e^{\rm cc}=\theta_e^{\rm max}$, i.e.,
where  meniscus depinning meets capillary condensation. Meniscus depinning only occurs in the range $\theta_p<\theta<\pi/4$ since for smaller contact
angles the CL is metastable. As the aspect ratio approaches unity both $\theta_p$ and $\theta_0$ approach $\pi/4$ (from different sides) and the
lines of type I condensation and meniscus depinning vanish.

{\bf Short capillaries} (Fig.~2c). If the aspect ratio $a>1$ only type II capillary condensation occurs up to the corner filling phase boundary with
$\theta_0=\pi/4$ beyond which capillary condensation is suppressed.

An alternative way of representing these macroscopic predictions is in terms of phase diagrams which are qualitatively different for the regimes
corresponding to complete corner filling ($\theta<\pi/4$) and partial corner filling ($\theta>\pi/4$), see Fig.~\ref{dp-a} and the caption for
details.

Some of these macroscopic predictions are slightly modified when we allow for thermal fluctuations. Since the capillary is pseudo one-dimensional all
capillary condensation transitions are rounded occurring over a pressure range $\Delta p_{\rm cc} \propto \exp(-\beta \gamma L H)$ where the factor
$\gamma LH$ is the approximate free-energy cost of phase separating the CG and CL along the capillary \cite{privman}. Such rounding is only of
significance in the near vicinity of the (pseudo) capillary critical temperature which itself occurs when the smallest of the dimensions $L$ or $H$
is of order the bulk correlation length. The meniscus depinning transition is also rounded. We consider the case of complete wetting first where the
rounding is largest. At a macroscopic level meniscus depinning occurs when $R=L$, i.e., when a quarter circular meniscus just fits into the open ends
of the capillary. However, this ignores the presence of the complete wetting layers along the bottom and vertical walls which are characterized by a
thickness $\ell_\pi$ and also a parallel correlation length $\xi_\parallel$ arising from thermal interfacial fluctuations \cite{dietrich}. These
length-scales soften the effective slit width, with depinning occurring when $R\approx L-\ell_\pi\pm \xi_\parallel$. Allowing for the pressure
dependence of the parallel correlation length, $\xi_\parallel \approx \delta p^{-\nu_{\parallel}^{\rm co}}$, implies that for complete wetting  the
meniscus depinning transition is rounded over the pressure range $ \Delta p_{\rm md}\propto L^{\nu_\parallel^{\rm co}-2}$ and the type I/II crossover
over the aspect ratio range $\Delta a_p\propto  L^{\nu_\parallel^{\rm co}-1}$. For systems with dispersion forces, $\nu_\parallel^{\rm co}=2/3$
\cite{lipowsky85}, implying that $\Delta p_{\rm md} \propto L^{-\frac{4}{3}}$ and $\Delta a_{\rm p} \propto L^{-\frac{1}{3}}$. Similar considerations
apply for partial wetting, however, in this case $\xi_\parallel$ remains finite leading to the universal finite-size scaling predictions $\Delta
p_{\rm md}\propto L^{-2}$ and $\Delta a_{\rm p} \propto L^{-1}$. The macroscopic predictions are also modified slightly if the corner filling
transition at $\theta=\pi/4$ is continuous. Macroscopically, the loci of the meniscus depinning transitions (for $a<1$) and line of type II
condensation (for $a>1$) end exactly at $\theta=\pi/4$. If the right-angle corners show continuous corner filling, however, these transitions end
when $\theta-\pi/4 \propto L^{-1/\beta_w}$. Here, $\beta_w$ is the critical exponent describing the thickness, $\ell_w \propto (\theta
-\frac{\pi}{4})^{-\beta_w}$, of the adsorbed layer of liquid at a right-angle corner at $p=p_{\rm sat}$ \cite{parry2000}, i.e., a meniscus must be
present whenever  the mesoscopic thickness of the adsorbed liquid is greater than the slit width. For systems with dispersion forces this implies
that meniscus depinning slightly extends into the partial filling regime until $\theta-\frac{\pi}{4} \propto L^{-2}$.

\begin{figure*}

\includegraphics[width=6.2cm]{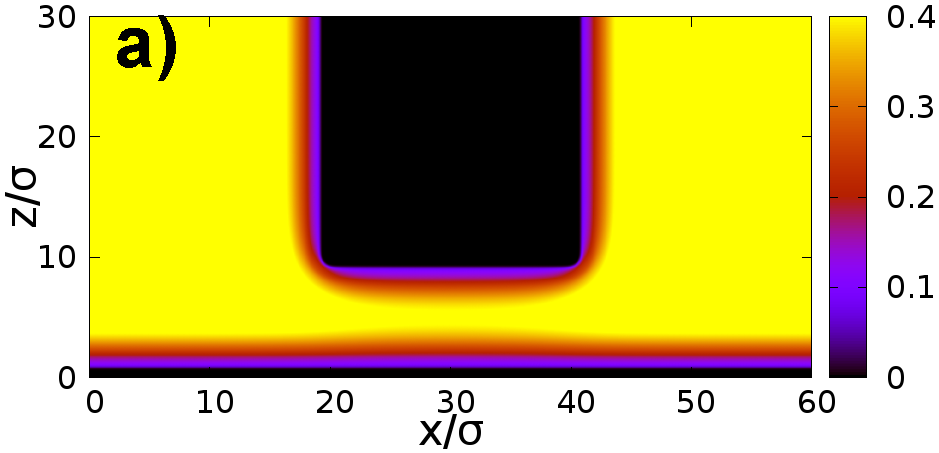} \hspace*{0.5cm} \includegraphics[width=6.2cm]{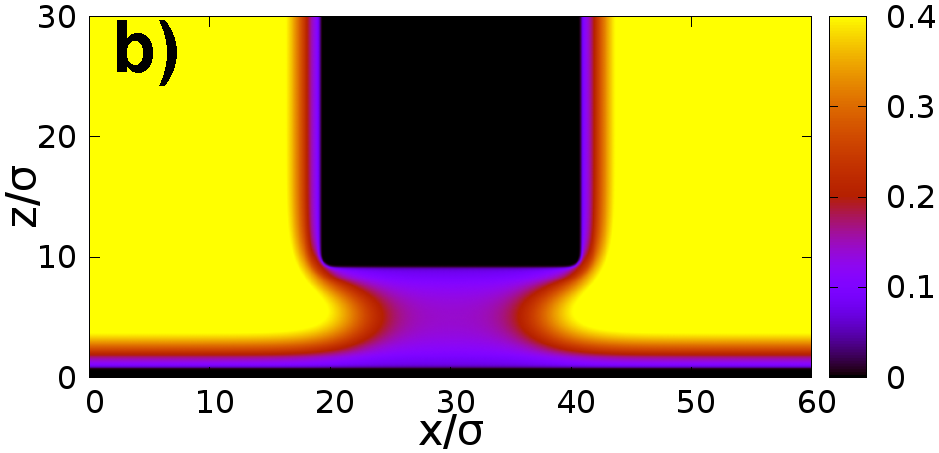}

\vspace*{0.3cm}

\includegraphics[width=6.2cm]{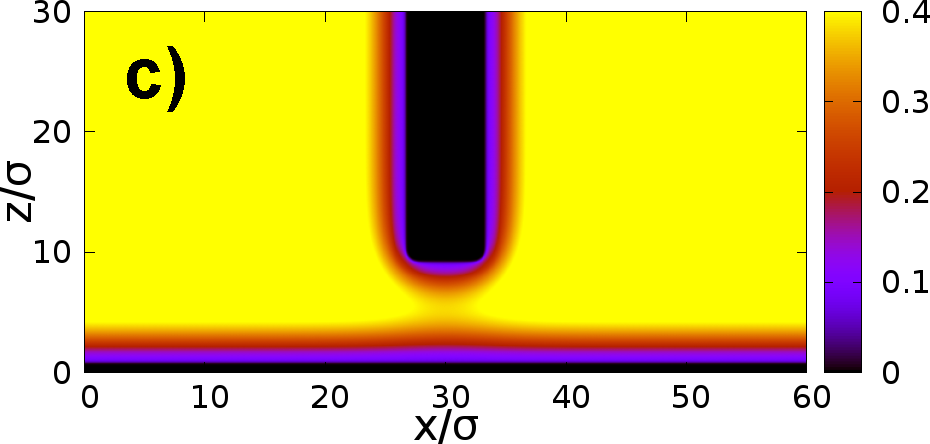} \hspace*{0.5cm} \includegraphics[width=6.2cm]{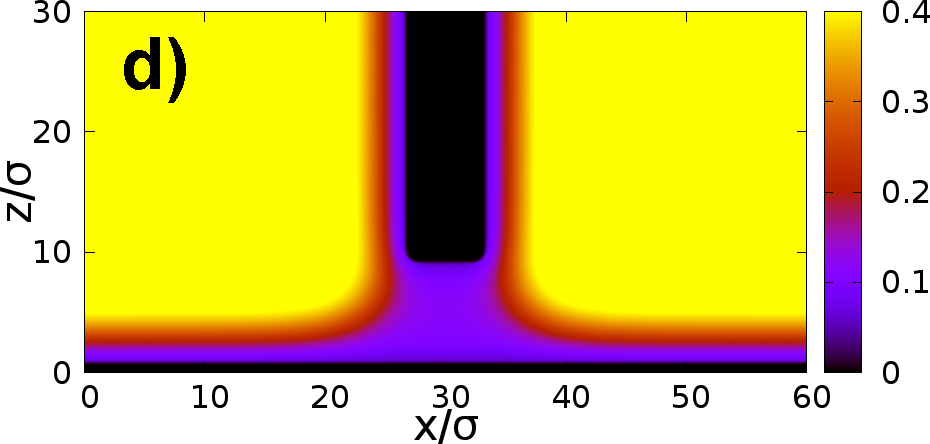}

\caption{Type I and type II condensation in a microscopic slit with repulsive walls and width $L=10\,\sigma$. The plots a) and b) show the coexisting
CL and CG phases for aspect ratio $a=1/2$, while c) and d)  are the coexisting CL and CG phases for $a=1$.} \label{dens_profs_H20}
\end{figure*}

\begin{figure}
\includegraphics[width=8cm]{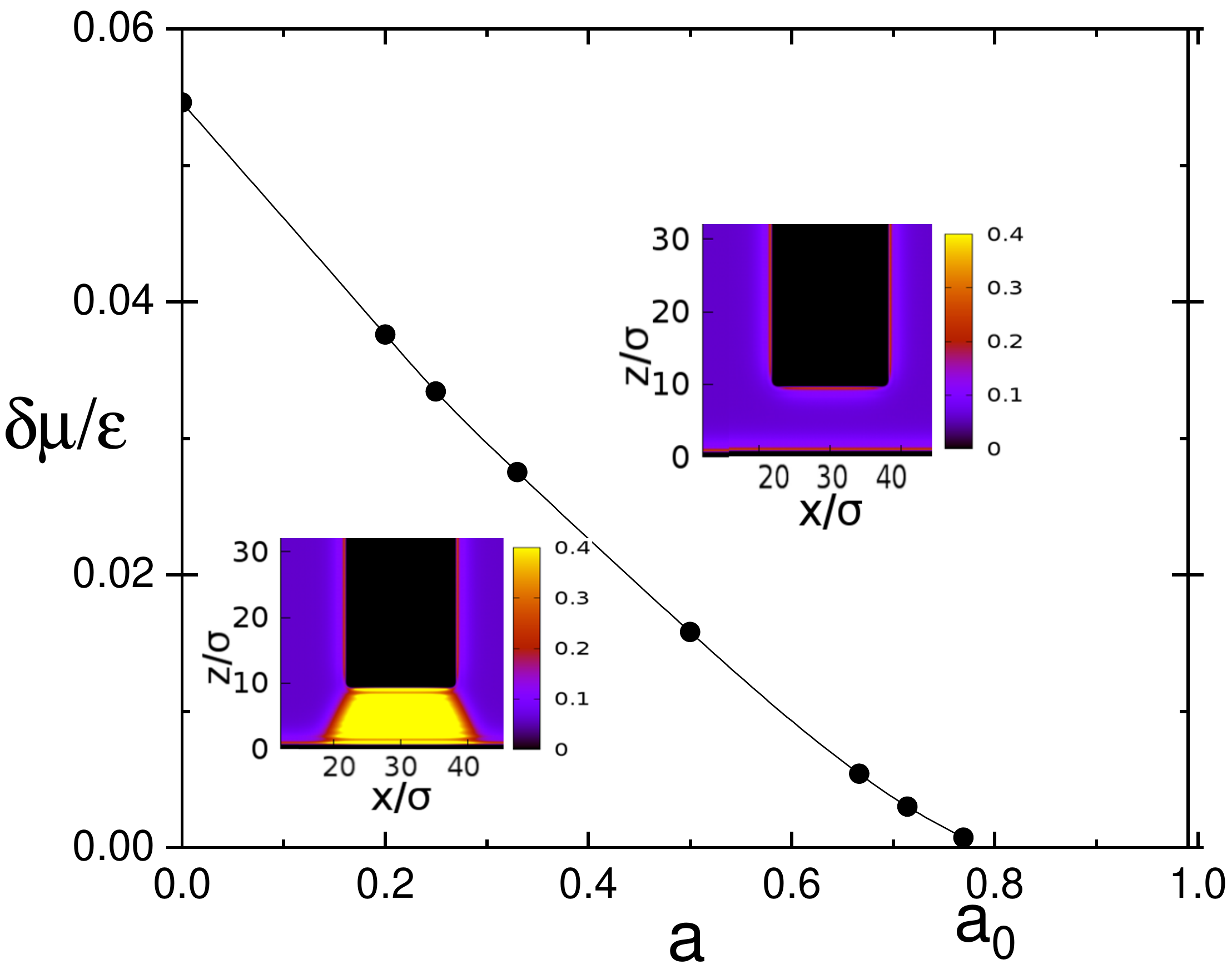}
\caption{Phase diagram for partial corner filling, with $\theta\approx 53\degree$ showing a line of type I capillary condensation ending at bulk
saturation when the aspect ratio $a_0\approx 0.78$, close to the macroscopic prediction $a_0\approx 0.75$. Here $\delta\mu/\varepsilon$ is a measure
of the chemical potential deviation from saturation measured in units of fluid-fluid potential. Coexisting density profiles for $a=2/3$ are shown.}
\label{fig_dft_t12}
\end{figure}

We have compared our predictions with a microscopic density functional theory (DFT) model  which  allows us to study these phenomena at the molecular
scale \cite{evans79}. We begin by showing that for complete wetting, the capillary condensation is type I and type II for small and large aspect
ratios, respectively.  We employ the same DFT model that we have used recently \cite{laska20} which combines Rosenfeld's fundamental measure theory
\cite{ros} describing accurately any packing effects, with a mean-field treatment of the attractive part of the inter-atomic interaction modelled by
a truncated Lennard-Jones (LJ) potential. See the Supplementary Material for details  \cite{sm}. Actually, we flip the scenario and consider walls
which have a purely long-ranged repulsive component, which ensures that they are completely dry with contact angle $\theta=\pi$, focussing on the
character of the capillary evaporation as the pressure is \emph{reduced} to $p_{\rm sat}$ (that is the roles played by the CG and CL phases are
simply reversed). We have determined the line of capillary condensation over a wide range of the aspect ratio for a microscopic slit separation
$L=10\,\sigma$, with $\sigma$ the molecular diameter \cite{sm}. Figs.~\ref{dens_profs_H20} a) and b) show the coexisting CL and CG phases for aspect
ratio $a=1/2$, illustrating type I condensation, while Figs.~\ref{dens_profs_H20} c) and d) show the coexisting phases for $a=1$, illustrating type
II condensation.

Finally, we show that for $\theta>\pi/4$ condensation is only of type I and is suppressed for aspect ratios $a>a_0$, which we compare with the
theoretical prediction $a_0=\cot\theta$. We add an attractive part to the substrate-fluid potential in order to decrease the contact angle assuming
the walls are made of atoms interacting with the fluid via a full LJ potential. We set the temperature  $T=0.85\,T_c$, for which the  contact angle
$\theta\approx53\degree$ \cite{bridge}. As predicted, the phase diagram shows only a line of type I capillary condensation which terminates at
$a_0\approx0.78$, which is extremely close to the macroscopic prediction $a_0\approx0.75$ (Fig.~\ref{fig_dft_t12}). Representative density profiles
of the coexisting states (for $H=15\,\sigma$), for which $a=2/3$, are also shown and illustrate how the meniscus pinning is mimicking the properties
of condensation between two walls with opposing wetting properties, i.e, $\theta_e^{\rm cc}\approx\pi-\theta$.

In summary, we have shown that in an open slit geometry the capillary condensation may occur in two different ways involving pinned or unpinned
menisci separated by a continuous meniscus depinning transition. The phase boundaries are determined by the values of an edge contact angle somewhat
analogous to how the Young contact angle determines the phase boundary for wetting and filling transitions. The resulting phase diagrams fall into
one of three possible universal classes depending on the slit aspect ratio. The richness of the possible phase behaviour emerges from the interplay
between different interfacial phenomena and are connected to the fluctuation theory of fundamental surface phase transitions. The distinction between
type I and type II condensation and the presence of meniscus depinning transitions will occur in other geometries which are certainly experimentally
accessible; for example, when a vertical cylinder is brought into close contact with a flat surface. The rounding of the meniscus transition
considered here arises due to the thermal fluctuations of the adsorbed wetting layers and occurs for even perfectly sharp geometries. It would also
be interesting to understand how surface roughness affects the edge contact angle and the meniscus depinning transition, which may well connect with
the phenomena of contact angle hysteresis. Including gravity may also introduce interesting new effects associated with capillary emptying
transitions \cite{dirk1, dirk2}. Finally, the equilibrium phase transitions considered here are also a pre-requisite for understanding the dynamics
of meniscus depinning which may be studied, for example, using dynamical DFT or simulation methods similar to those described in \cite{trobo}.

This work was financially supported by the Czech Science Foundation, Project No. GA 20-14547S.

\end{document}